\documentclass[twocolumn,prl,showpacs,showkeys]{revtex4}
\usepackage{hyperref}
\usepackage{amsmath}
\usepackage{bm}
\usepackage{graphicx}

\begin{document}

\title{Random Delays and the Synchronization of Chaotic Maps}

\author{C. Masoller$^{1,2}$ and A. C.  Mart\'{\i}$^1$}
\affiliation{$^1$Instituto de F\'{\i}sica, Facultad de Ciencias,
  Universidad de la Rep\'ublica,
  Igu\'a 4225, 11400 Montevideo, Uruguay \\
  $^2$Departament de Fisica i Enginyeria Nuclear, Universitat
  Politecnica de Catalunya, Colom 11, E-08222 Terrassa, Spain }

\date{\today}

\begin{abstract}
  We investigate the dynamics of an array of logistic maps
  coupled with random delay times. We report that for adequate
  coupling strength the array is able to synchronize, in spite of the
  random delays. Specifically, we find that the synchronized state is a
  homogeneous steady-state, where the chaotic dynamics of the
  individual maps is suppressed. This differs drastically from the synchronization with instantaneous and
  fixed-delay coupling, as in those cases the dynamics is chaotic. Also in contrast with the instantaneous and
  fixed-delay cases, the synchronization does not dependent on the connection topology, depends only on the
  average number of links per node. We find a scaling law that relates
  the distance to synchronization with the randomness of the delays. We also carry out a statistical
  linear stability analysis that confirms the numerical results and provides a better understanding of the nontrivial
  roles of random delayed interactions.
\end{abstract}

\maketitle

A system composed of many nonlinear interacting units often forms a
complex system with new emergent properties that are not held by the
individual units. Such systems describe a wide variety of phenomena in
biology, physics, and chemistry. The emergent property is usually
synchronous oscillations. In biology examples are the synchronized
activity in pacemaker heart cells, the cicardian rhythms that control
24h daily activity, and the flashing on-and-off in unison of
populations of fireflies. Nonbiological examples include synchronized
oscillations in arrays of electrochemical reactions, in laser arrays,
and in Josephson junction arrays \cite{reviews}.

The effect of time-delayed interactions, which arise from a
realistic consideration of finite communication times, is a key
issue that has received considerable attention. The first systematic
investigation of time-delayed coupling was done by Schuster and
Wagner \cite{schuster_PTP_1989}, who studied two coupled phase
oscillators and found multistability of synchronized solutions.
Since then, delayed interactions have been studied in the context of
linear systems \cite{jirsa_PRL_2004}, phase oscillators
\cite{phase_osc}, limit-cycle oscillators \cite{limit_cycl}, coupled
maps \cite{maps,atay_PRL_2004}, neuronal
\cite{neuronal,misha,longtin} and laser \cite{laser} systems. Most
studies have assumed that all the interactions occur with the same
delay time (only few have consider non-uniform delays
\cite{longtin,zanette_PRE_2000,marti}). However, actual delays in
real extended systems are not necessarily the same for all the
elements of the system, they might be distant-dependent or randomly
distributed. In populations of spatially separated neurons, the
synaptic communications between them, which depend on the
propagation of action potentials over appreciable distances, involve
distributed delays. In computer networks, random delays arise from
queueing times and propagation times. In epidemic dynamics,
migrations of geographically spread populations lead to different
delays which influence the transmission of diseases.

While it is well-known that oscillators which interact with
different delay times can synchronize (an example is the synchrony
arising in different neuronal groups of the brain which might lead
to both, epilepsy and Parkinson disease), the mechanism by which
this synchrony arises and the influence of the random communication
times remains poorly understood. The focus of this letter is to
investigate the influence of such random delays in the
synchronization of a simple model of coupled chaotic oscillators. We
consider an ensemble of logistic maps and show that, in spite of the
random delays, for adequate coupling strength the array is able to
synchronize. Surprisingly, in the synchronized state the chaotic
dynamics of the individual maps is suppressed: the maps are in a
steady-state, which is unstable for the uncoupled maps. This is in
sharp contrast with the cases of instantaneous and fixed-delay
coupling, as in those cases the dynamics of the array is chaotic. By
studying the transition from chaotic synchronization to steady-state
synchronization as the randomness of the delays increases we
discover a scaling law that relates the distance to synchronization
with the randomness of the delays. We also investigate the influence
of the array topology and find that steady-state synchronization
depends on the average number of links per node but not on the array
architecture. This is also in contrast with the instantaneous and
fixed-delay cases, as in those cases the synchronization depends
also on the connection topology \cite{atay_PRL_2004}. Finally, we
present a statistical linear stability analysis that demonstrates
the stability of the solution found numerically.

We consider the following ensemble of $N$ coupled maps:

\begin{equation}
\label{mapa}
x_i(t+1)= (1-\epsilon) f[x_i(t)] +
\frac{\epsilon} {b_i} \sum_{j=1}^N \eta_{ij} f[x_j(t-\tau_{ij})]
\end{equation}

Here $t$ is a discrete time index, $i$ is a discrete spatial index
($i=1\dots N$), $f(x)=ax(1-x)$ is the logistic map, the matrix
$\eta=(\eta_{ij})$ defines the connectivity of the array:
$\eta_{ij}=\eta_{ji}=1$ if there is a link between the $i$th and
$j$th nodes, and zero otherwise. $\epsilon$ is the coupling strength
and $\tau_{ij}$ is the delay time in the interaction between the
$i$th and $j$th nodes (the delay times $\tau_{ij}$ and $\tau_{ji}$
need not be equal). The sum in Eq.(\ref{mapa}) runs over the $b_i$
nodes which are coupled to the $i$th node ($b_i = \sum_j
\eta_{ij}$). The normalized pre-factor $1/b_i$ means that each map
receives the same total input from its neighbours.

First we note that the homogeneous steady-state
$x_i(t)=x_j(t)=x_0$,
$\forall$ $i$, $j$, $t$, where $x_0$ is a fixed point of the
uncoupled map, $x_0=f(x_0)$, is a solution of Eq.(\ref{mapa})
regardless of the delays and of the connectivity of the array.

Next, let us present some results of simulations that show that this
state, with $x_0$ being the nontrivial fixed point, $x_0=1-1/a$, can
be a stable solution for adequate coupling and random enough delays.
The simulations were done choosing an initial configuration,
$x_i(0)$ random in [0,1], and letting the array evolve initially
without coupling [in the first time interval $0<t<\max(\tau_{ij})$].
We present results for $a=4$, corresponding to fully developed chaos
of the individual maps, but we have found similar results for other
values of $a$. We illustrate our findings using the small-world
topology \cite{nw_1999}, but we have found similar results for
random and regular topologies \cite{nota1}, as discussed below.

\begin{figure}
  \center
  \resizebox{1.0\columnwidth}{!}{\includegraphics{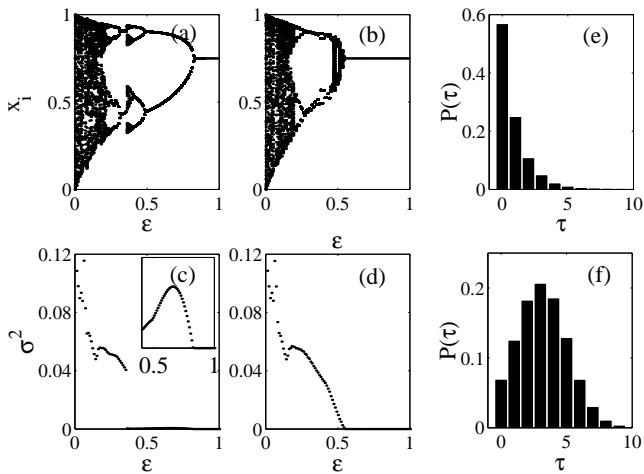}}
  \caption{$x_i$ vs. $\epsilon$ (a) and (b);
    $\sigma^2$ vs. $\epsilon$, (c) and (d).  In Figs. 1(a) and 1(c)
    the delays are distributed exponentially ($\tau_0=0$,
    $c=1.2$, see text); the distribution is shown in Fig. 1(e).
    In Figs. 1(b) and 1(d) the delays are Gaussian distributed ($\tau_{0}=3$,
    $c=2$, see text); the distribution
    is shown in Fig. 1(f). The inset in Fig. 1(d) shows with detail
    the transition to synchronization: $\sigma^2$ decreases abruptly
    at $\epsilon \sim 0.4$, and is zero for $\epsilon > 0.8$.
    Parameters are: $N=500$, $a=4$, and $p=0.3$.}
\end{figure}

With both, either randomly distributed or fixed delay times, if the
coupling is large enough the array synchronizes in a spatially
homogeneous state: $x_i(t)=x_j(t)$ $\forall$ $i,j$. Figures 1 and 2
display the transition to synchronization as $\epsilon$ increases.
At each value of $\epsilon$, 100 iterates of an element of the array
are plotted after transients. To do the bifurcation diagrams we
varied only $\epsilon$; the connectivity of the array ($\eta_{ij}$),
the delays ($\tau_{ij}$), and the initial configuration [$x_i(0)$]
are the same for all values of $\epsilon$. Figure 1(a) displays
results for delays that are exponentially distributed, Fig. 1(b) for
Gaussian distributed, Fig. 2(a) for zero delay, and Fig. 2(b) for
constant delays. It can be observed that for small $\epsilon$ the
four bifurcation diagrams are similar; however, for large $\epsilon$
they differ drastically: $x_i$ is constant in Figs. 1(a) and 1(b),
$x_i=x_0=1-1/a$, while $x_i$ varies within [0,1] in Figs. 2(a) and
2(b).

\begin{figure}
\center
\resizebox{0.8\columnwidth}{!}{\includegraphics{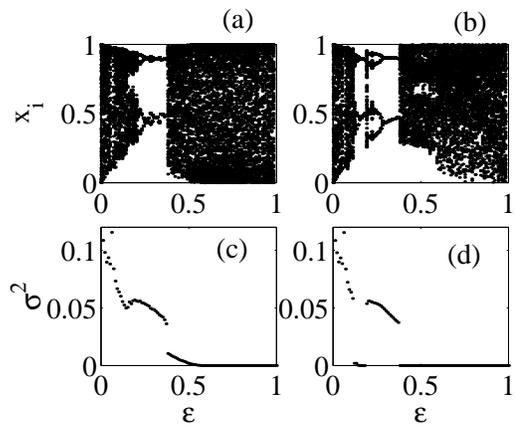}}
\caption{$x_i$ vs. $\epsilon$ (a) and (b); $\sigma^2$ vs. $\epsilon$
(c) and (d). In (a),(c) $\tau_{ij}=0$ $\forall$ $i$,$j$; in (b),(d)
$\tau_{ij}=3$ $\forall$ $i$,$j$. All other parameters as in Fig. 1.}
\end{figure}

To characterize the transition to synchronization we use the indicator
$\sigma^2 = 1/N<\sum_i [x_i(t)-<x>]^2>_t$,
where $<.>$ denotes an average over the elements of the array and
$<.>_t$ denotes an average over time. Figures 1(c), 1(d), 2(c) and
2(d) display $\sigma^2$ vs. $\epsilon$ for the bifurcation diagrams
discussed above. It can be observed that for large $\epsilon$ there
is inphase synchronization in the four cases [$x_i(t)=x_j(t)$ and
$\sigma^2=0$]; however, we remark that an inspection of the
time-dependent dynamics reveals that for randomly distributed delays
the maps are in a steady-state, while for fixed delays the maps
evolve chaotically. It can also be observed that the four plots
$\sigma^2$ vs. $\epsilon$ are similar for small $\epsilon$ (in Fig.
2(d) the array synchronizes also in a window of small $\epsilon$;
this occurs for odd delays and was reported in
\cite{atay_PRL_2004}).

Let us now investigate the transition from chaotic synchronization
(for fixed delays) to steady-state synchronization (for random
delays) by introducing a disorder parameter $c$ that allows varying
the delays from constant to distributed values. Specifically we
consider

i) $\tau_{ij} = \tau_0 + \mathrm{Near} (c \xi)$, where $\xi$ is
Gaussian distributed with zero mean and standard deviation one and
$Near$ denotes the nearest integer. The delays are constant
($\tau_{ij}=\tau_0$) for $c=0$ and are Gaussian distributed around
$\tau_0$ for $c\ne 0$ [depending on $\tau_0$ and $c$ the
distribution of delays has to be truncated to avoid negative delays,
see Fig. 1(f)].

ii) $\tau_{ij} = \tau_0 + \mathrm{Int} (c \xi)$, where $\xi$ is
exponentially distributed, positive, with unit mean and $Int$
denotes integer. The delays are constant ($\tau_{ij}=\tau_0$) for
$c=0$ and are exponentially distributed, decaying from $\tau_0$ for
$c\ne 0$.

Figure 3(a) displays the transition between the two synchronization
regimes as the randomness of the delays increases. We plot
$\sigma^2$ vs. $c$ for different delay time distributions and
$\epsilon$ large enough such that the array synchronizes for $c=0$.
It can be observed that the array desynchronizes as $c$ increases
and the delays become different from each other. There is a range of
values of $c$ such that the delays are not random enough to induce
steady-state synchronization; however, for $c$ large enough the
array synchronizes again and $\sigma^2=0$.

To investigate this transition we considered a normalized disorder parameter
$c^*=D_\tau/<\tau>$, where $<\tau>$ is the average delay and $D_\tau$ is the standard deviation of the delay distribution. By plotting $\sigma^2$ vs.
$c^*$ [Figs. 3(b)] we uncover a scaling law: as $<\tau>$ increases the curves collapse into curves of similar shape, and the transition to steady-state synchronization occurs for $c^*\sim 0.5$.

\begin{figure}
\center
\resizebox{1.0\columnwidth}{!}{\includegraphics{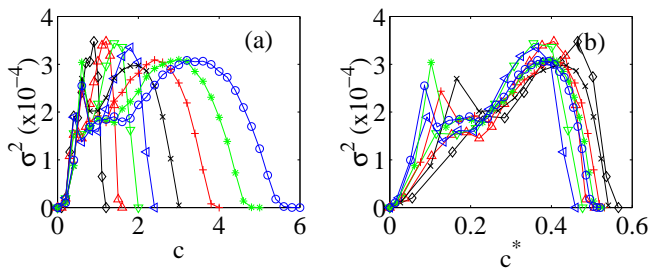}}
\caption{ (a) $\sigma^2$ vs. $c$ and (b) $\sigma^2$ vs. $c^*$. The
delays are Gaussian distributed with $\tau_0=2$ ($\diamond$), 3
($\triangle$), 4 ($\nabla$), 5 ($\triangleleft$) ; exponentially
distributed with $\tau_0=3$ ($\times$), 4 ($+$), 5 ($\ast$), 6
($\circ$). $\epsilon=1$, all other parameters are as in Fig. 1.}
\end{figure}

The value of the disorder parameter above which the arrays
synchronizes in the steady-state depends on the coupling strength.
Figure 4(a) displays the synchronization region in the parameter
space $(\epsilon,c)$. To determine the synchronization region we did
simulations with different initial conditions, array connectivities,
and delay time realizations: the black indicates parameters for
which the array synchronized in all the simulations, while the white
indicates parameters for which the array did not synchronize in any
of the simulations. The grey region in the boundary of the
synchronization region indicates that there are some initial
conditions and/or  realizations of $\eta_{ij}$ and $\tau_{ij}$ for
which the array did not synchronize. Two different synchronization
regions can be clearly distinguished: for $c=0$ and for $c>0.5$. The
former corresponds to chaotic synchronization for fixed delays, and
the latter, to steady-state synchronization for distributed delays.

Atay et al. \cite{atay_PRL_2004} have recently shown that with fixed
delays the synchronization depends on the array architecture: with
the same total number of links, a random network exhibits better
synchronization properties than a regular network. We investigated
this issue in the case of distributed delays and found that the
synchronization depends on the average number of links per node,
$<b_i>$, but not on the architecture. We did simulations with arrays
of different topologies \cite{nota1} and found that the transition
to synchronization is similar for all the arrays provided that they
have the same $<b_i>$. Figure 5(a) displays the transition to
synchronization as $\epsilon$ increases for small-world and regular
arrays with distributed delays, and for comparison, Fig. 5(b)
displays results for the same arrays with fixed-delays. It can be
observed that for distributed delays the transition to
synchronization is independent of the array topology; however, it
depends on the connectivity: the larger the number of mean links per
node, the lower the coupling strength needed synchronize. In
contrast, for fixed delays the synchronization depends not only on
the connectivity but also on the architecture: for $\epsilon$ large enough
the arrays that have small-world topologies synchronize, but those
that have regular topologies do not, in agreement with the results
of \cite{atay_PRL_2004}.

\begin{figure}
\center
\resizebox{1.0\columnwidth}{!}{\includegraphics{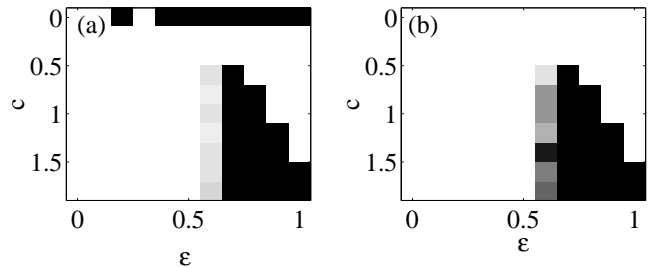}}
\caption{ (a) Synchronization region determined numerically: black
represents parameters where $\sigma^2<10^{-7}$ for all the
realizations of $x_i(0)$, $\eta_{ij}$ and $\tau_{ij}$. (b)
Synchronization region determined from the stability analysis: black
represents parameters where the $|\lambda_{max}|<1$ for all the
realizations of $\eta_{ij}$ and $\tau_{ij}$. The delays are Gaussian
distributed with $\tau_{0}=3$, $N=100$, all other parameters are as
in Fig. 1. }
\end{figure}

\begin{figure}
\center
\resizebox{1.0\columnwidth}{!}{\includegraphics{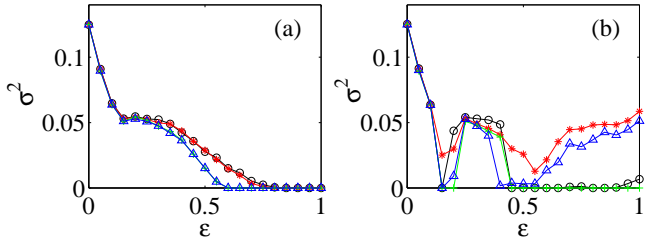}}
\caption{Influence of the architecture and the connectivity of the
array for (a) Gaussian distributed delays ($\tau_0=3$, $c=2$) and
(b) fixed delays ($\tau_{ij}=3$ $\forall$ $i$,$j$). Small-world
topology with and an average of 10 links per node ($\circ$), 50
links per node (+); regular topology with 10 links per node (*); 50
links per node ($\triangle$). All other parameters are as in Fig.
1.}
\end{figure}

Finally, let us assess the stability of the steady-state
synchronized behavior found numerically by performing a statistical
linear stability analysis. The delayed map Eq.~(\ref{mapa}) can be
written in non-delayed form by the introduction of a set of
auxiliary variables \cite{marti}, $y_{im}(t)=x_i(t-m)$, where $1\leq
i \leq N$ and $0 \leq m \leq M$ with $M=\max(\tau_{ij})$. In terms
of these new $N(M+1)$ variables Eq.~(\ref{mapa}) becomes

\begin{eqnarray}
\label{nueva}
y_{im}(t+1)=
\nonumber
\\
\begin{cases}
              y_{i,m-1}(t)& \text{if $m \ne 0$},\\
              (1-\epsilon)f[y_{i0}(t)]+
              \sum_{j=1}^N \epsilon_{ij} f[y_{j,\tau_{ij}}(t)] &
                                        \text{if $m=0$}.
              \end{cases}
\end{eqnarray}

Next we define a vector of $N(M+1)$ components containing
information about the present and past states of the array (for
$i=1$ to $N$ and for time $t$ to $t-M$):
$Y=(y_{10},y_{20},\dots,y_{N0};y_{11},y_{21}, \dots, y_{N1};\dots
;y_{1M}, y_{2M}\dots,y_{NM})$. Eq.~(\ref{nueva}) can be re-written
as $Y_i(t+1) = F_i(Y_1(t)\dots Y_{N(M+1)}(t))$ and the synchronized
state can be re-written as $ Y^* = (x_0 \dots x_0).$

Linearizing the equations of motion near $Y^*$ gives $\delta Y_i(t)
= \sum_j f_{ij} \delta Y_j(t)$, where the matrix ${\bf F}=(f_{ij})$
with $f_{ij}=\partial F_i/\partial Y_j$ can be cast as a set of
$(M+1)^2$ blocks of dimension $N\times N$.  We denote these blocks
as $\mathcal{F}_{kl}$, with $k,l=0,...M$. The blocks
$\mathcal{F}_{kl}$ with $k>0$ have all components equal to $0$,
except the blocks $\mathcal{F}_{k+1,k}$ which are $N \times N$
identity matrices. The elements of the blocks $\mathcal{F}_{kl}$
with $k=0$ can be considered as composed by two parts: (i)
$(1-\epsilon)f^\prime(x_0)$, which arise from the non-delayed term
in Eq.~(\ref{mapa}) and appear in the diagonal of
$\mathcal{F}_{00}$. (ii) $(\epsilon/b_i)f^\prime(x_0)$, which
represent the contribution of a link and due to the random delays
appear in random positions of the blocks.  Each link contributes
with two terms of the form $(\epsilon/b_i)f^\prime(x_0)$. There is a
total of $2\sum_i b_i$ terms $(\epsilon/b_i)f^\prime(x_0)$ are which
are distributed randomly in the blocks $\mathcal{F}_{0l}$. More
precisely, they are located at the positions $(i,j)$ and $(j,i)$ in
the blocks $\mathcal{F}_{0\tau_{ij}}$ and $\mathcal{F}_{0\tau_{ji}}$
respectively.  The rest of the elements are zero.

We calculated the eigenvalues of ${\bf F}$ for different
realizations of the connectivity and delay time distributions. The
results are displayed in Fig. 4(b), where the black region indicates
parameters where the maximum eigenvalue of $\bf{F}$,
$\lambda_{max}$, has modulus less than 1 for all $\tau_{ij}$ and
$\eta_{ij}$ realizations, and the white region indicates parameters
where $\lambda_{max}\ge 1$ for all $\tau_{ij}$ and $\eta_{ij}$
realizations. A very good agreement with the synchronization region
determined numerically can be observed [the black region for $c=0$
observed in Fig. 4(a) does not appear in Fig. 4(b) because in this
region the synchronized dynamics is chaotic].

To conclude, we studied the dynamics of an ensemble of chaotic maps
which interact with random delay times, and found that for adequate
coupling strength the array synchronizes with all the maps in a
steady-state. The synchronization is independent of the architecture
of the array, but depends on the average number of links per node.
We confirmed the numerical results by performing a statistical
linear stability analysis. Our findings provide another example of
the nontrivial action of inhomogeneities and disorder in coupled
nonlinear systems; in our case, the presence of large enough
disorder in the interaction times plays a constructive role,
suppressing the chaotic dynamics of the individual units. We
speculate that the "chaos-suppression by random delays" reported
here might yield light into explaining the stable operation of many
complex systems composed by nonlinear units which interact with each
other with random communication times.


\begin{thebibliography}{99}
\bibitem{nota1} To built a small-world topology we begin with a circular chain of $N$ nodes and add with probability $p$ a link between two nodes ($p=0$ corresponds to a circular array with
nearest-neighbour coupling, $p=1$ corresponds to a globally coupled
array); to built a random topology we begin with $N$ nodes and add
with probability $C$ ($0\le C \le 1$) a link between two nodes; to
built a regular topology we begin with $N$ nodes and couple each
node with its nearest $K$ neighbors.
\bibitem{nota2} Unfortunatelly, we were not able to investigate system size
effects because of the large memory requirements needed to simulate
large arrays with delayed coupling. To the best of the numerical
resolution we have not observed significant differences in the
synchronization of arrays with small-world, random and regular
topologies, as long as the average number of links per node,
$<b_i>$, is equal. The results of the stability analysis confirm
this numerical prediction.
\end{thebibliography}
\end{document}